# Photon shot-noise-limited Rydberg-EIT electrometry

GYEONGMIN JU,[1] YE JIN YU,[1] HEEWOO KIM,[1] HANSOL JEONG,[1] JINHYUK BAE,[1] CHANGHOON BAEK,[1] AND HAN SEB MOON[1,2,3] *

[1]*Department of Physics, Pusan National University, Geumjeong-Gu, Busan 46241, Republic of Korea*
[2]*Quantum Sensors Research Center, Geumjeong-Gu, Busan 46241, Republic of Korea*
[3]*Quantum Science Technology Center, Pusan National University, Geumjeong-Gu, Busan 46241, Republic of Korea*
**hsmoon@pusan.ac.kr*

**Abstract:** Rydberg-atom electrometry is a core technique in the development of highly sensitive quantum electric-field sensors. Its sensitivity based on atom-photon interaction is typically limited by photon shot-noise (PSN) and spectral broadenings. Here, we experimentally demonstrate a near PSN-limited Rydberg electrometry from a $^{85}$Rb atomic vapor cell. By engineering atomic coherence through control of residual magnetic fields and laser frequency noise, we achieve the Rydberg electromagnetically induced transparency (EIT) with the narrow linewidth of 1.6 MHz, yielding an enhanced spectral slope for high-sensitivity Rydberg-EIT electrometry. Under optimized superheterodyne detection conditions, we obtain an electric-field sensitivity of 12.5(8) nV cm$^{-1}$ Hz$^{-1/2}$ at 37 GHz, in close agreement with the calculated PSN limit. These results provide direct experimental evidence of the high-sensitive quantum electrometry and establish a practical route toward quantum-noise-limited Rydberg electrometry.

## 1. Introduction

Rydberg-atom electrometry, a significant breakthrough in quantum sensing, provides a new paradigm for electromagnetic field detection [1-16]. It serves as a versatile platform for broadband, SI-traceable sensing, enabling direct measurement of radio-frequency (RF) and THz fields without conventional antennas or calibration procedures [6-20]. By exploiting the large electric dipole moments of highly excited Rydberg states, these high-sensitivity sensors provide strong and intrinsically calibrated coupling between electromagnetic fields and optical observables, spanning the GHz to THz range. These capabilities have enabled diverse applications in RF metrology, THz imaging, and a various quantum-enhanced electric field sensing [9-33].

The most common implementation relies on cascade-type electromagnetically induced transparency (EIT), where an RF field couples adjacent Rydberg states and modulates the optical response through Autler-Townes (AT) splitting or dispersive features [30-44]. Significant progress of Rydberg-atom sensors has been made in enhancing sensitivity through techniques such as superheterodyne detection, repumping schemes, and optimized operating conditions [24, 45-52]. Previous studies have identified key limiting factors, including laser frequency noise, residual amplitude noise, and magnetic-field-induced dephasing, all of which broaden the EIT resonance and reduce the achievable field responsivity [53-56]. These approaches have achieved the noise-equivalent electric field sensitivities on the order of tens of nV cm$^{-1}$ Hz$^{-1/2}$ in room-temperature vapor cells [5, 24, 57, 58]. However, it remains an open question whether room-temperature vapor-cell systems can approach the photon shot-noise (PSN) limit of optical readout, as they are hampered by Doppler broadening, transit-time effects, external magnetic field, and technical noises of lasers [58-60].

Realizing the PSN-limited sensitivity requires a simultaneous optimization of the EIT linewidth, signal transduction efficiency, and detection noise. In thermal vapor systems, this presents a significant challenge. There is a typical trade-off between narrowing the linewidth and maintaining sufficient optical signal strength, while technical noise sources can dominate the intrinsic quantum noise floor. Although theoretical analyses have established the PSN limit for Rydberg-based RF sensing, the fundamental sensitivity limit of Rydberg electrometers in atomic vapor cells has been not yet to be reached [61].

Here, we report Rydberg-EIT electrometry operating near the photon shot-noise limit in a room-temperature rubidium vapor cell. By compensating for residual magnetic fields and reducing laser intensity and frequency noise, we observe a Rydberg-EIT linewidth of 1.6 MHz, limited primarily by transit-time broadening. Combined with atomic superheterodyne detection and operation at the maximum spectral slope, this approach enables efficient conversion of microwave electric fields into optical signals. To reveal the heterodyne signal, we analyze the electric-field sensitivity at the optimal operating point. Under optimized conditions, for the first time, we achieve the electric-field sensitivity of close agreement with the theoretical PSN limit. More importantly, our work establishes a general framework for quantum-limited Rydberg electrometry by quantitatively identifying the interplay between EIT spectral responsivity and probe-photon shot noise. This approach clarifies the fundamental design principles governing vapor-cell Rydberg sensors and provides a scalable pathway toward compact quantum-limited microwave electrometry.

## 2. Experimental schematic

Figure 1(a) illustrates the energy-level diagram of the Rydberg electric-field sensor based on cascade-type EIT of $^{85}$Rb atoms and atomic superheterodyne detection. A weak probe field ( $\Omega_P$ ) drives the lower transition from the atomic ground state $|1\rangle$ ($5S_{1/2}$, F = 3) to the intermediate excited state $|2\rangle$ ($5P_{3/2}$, F′ = 4), while a counter-propagating coupling field ( $\Omega_C$ ) excites the atoms to a Rydberg level ($40D_{5/2}$). In the absence of microwave driving, the coherent interaction between the probe and coupling fields induces a narrow Rydberg-EIT resonance in the transmitted probe signal. Incident microwave electric fields ( $\Omega_{LO}$ and $\Omega_{SIG}$ ) couple the selected Rydberg state $|3\rangle$ ($40D_{5/2}$) to a nearby Rydberg level $|4\rangle$ ($39F_{7/2}$), effectively perturbing the dark-state condition. This interaction transduces the microwave-induced coherence into a measurable optical response.

In our configuration, a strong local oscillator (LO, $\Omega_{LO}$ ) dresses the Rydberg transition, while a weak signal field (SIG, $\Omega_{SIG}$ ) is detected through its beat note with the LO. This interaction generates an intermediate-frequency (IF) modulation in the probe transmission, which is read out by a low-noise balanced photodetector with quantum efficiency of ~95%. A microwave source drives the Rydberg transition ($40D_{5/2}$-$39F_{7/2}$) at ~37 GHz. In heterodyne operation, the applied field consists of a strong LO, yielding a Rabi frequency of $\Omega_{LO}/2\pi = 3.7\ \mathrm{MHz}$ , and a weak SIG offset by 40 kHz from the LO. Their frequency difference is down-converted by the atomic medium and appears as an IF modulation in the probe transmission.

This approach enables linear electric-field sensing at the point of maximum spectral slope and shifts the signal away from low-frequency technical noise, thereby improving the achievable noise floor. Achieving high sensitivity requires a narrow and stable Rydberg-EIT linewidth. In atomic vapor cells, the EIT linewidth is broadened by Doppler averaging, transit-time effects, environmental magnetic fields, and laser noise. In this work, we suppress magnetic fields and reduce laser intensity and frequency noise, stabilizing the EIT resonance and enabling operation close to the photon shot-noise (PSN) limit.

Figure 1(b) shows the experimental apparatus used to implement the compact Rydberg-EIT superheterodyne sensor. The sensing medium is a 50-mm-long room-temperature $^{85}$Rb vapor cell located at the center of the interaction region. A weak probe beam and a counter-propagating coupling beam overlap inside the cell to form the cascade-type EIT system. The probe and coupling powers are 6 μW and 30 mW, respectively, with a beam diameter of 0.15 mm, and their polarizations are linear and orthogonal. To facilitate balanced detection, the probe beam is split into two parallel paths using a beam displacer (BD). The transmitted probe fields are separated from the coupling beam using dichroic mirrors (DM) and detected by a balanced photodetector (BPD), which converts the optical modulation into an electrical IF signal. The counter-propagating geometry suppresses first-order Doppler shifts and ensures efficient mapping of the microwave-induced atomic response onto the probe transmission. Balanced detection further reduces intensity noise and enhances the signal-to-noise ratio.

To suppress environmental magnetic fields, the vapor cell is surrounded by three-axis compensation coils that minimize residual magnetic-field fluctuations. In addition, the frequency noise of both probe and coupling lasers is minimized by locking them to an ultra-low-expansion (ULE) reference cavity to prevent laser-induced broadening of the EIT resonance. Without this suppression, laser fluctuations would be directly converted into probe transmission noise and obscure the photon shot-noise floor. By reducing both magnetic-field fluctuations and laser noise, the Rydberg-EIT linewidth is limited primarily by transit-time broadening. Under these conditions, the sensor operates in a regime where the achievable sensitivity is determined by the intrinsic optical readout noise rather than by technical noise sources.

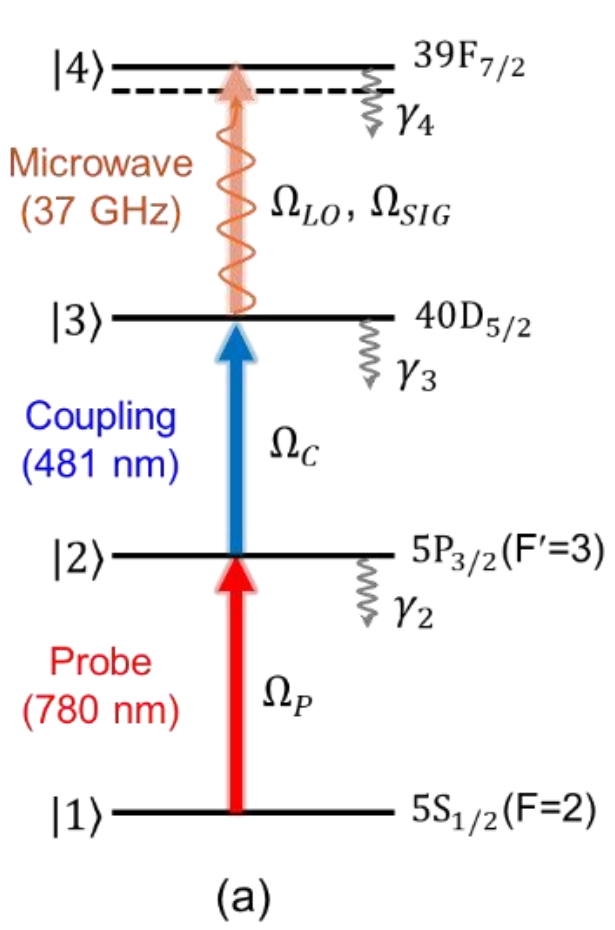

(a)

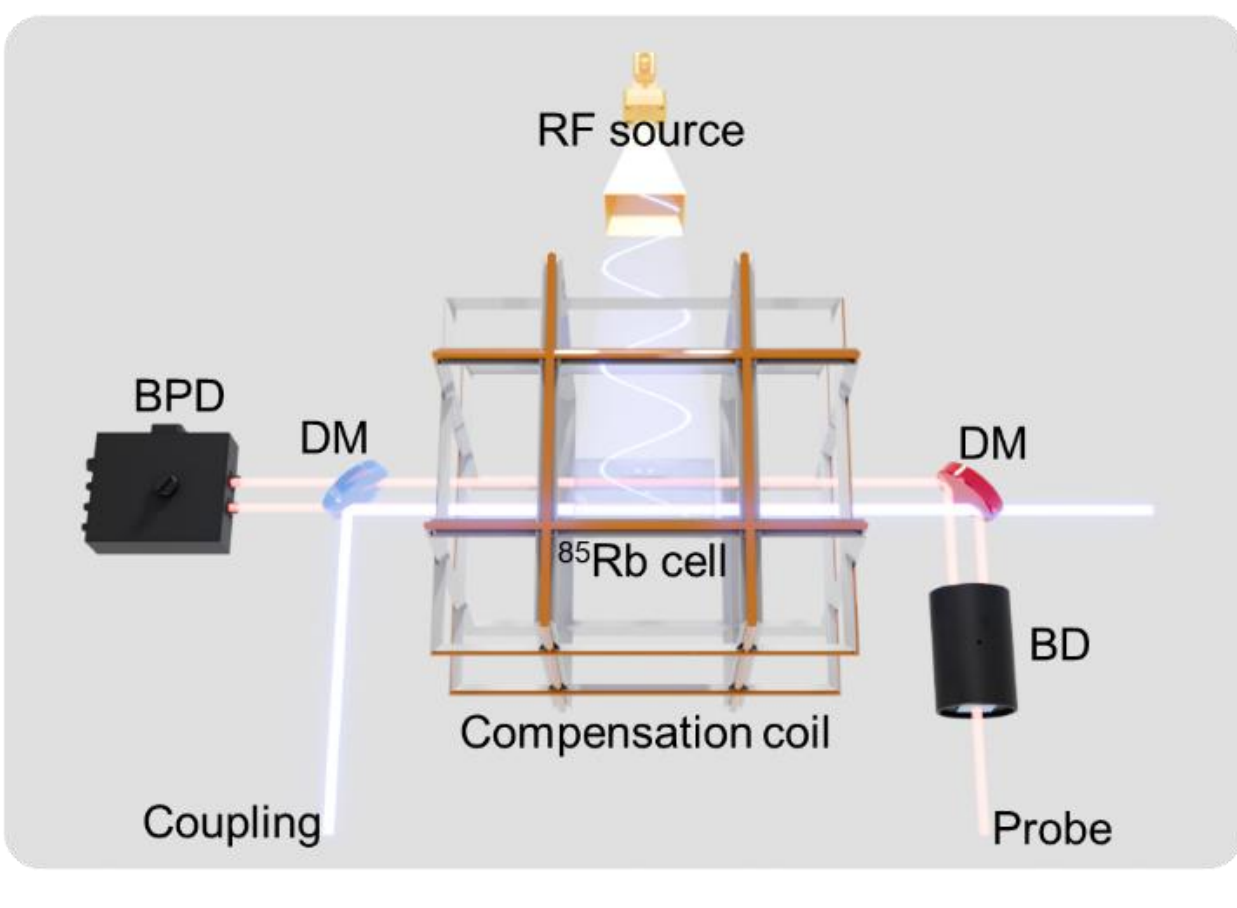

(b)

**Fig. 1. Energy diagram and experimental setup for Rydberg-EIT electrometry.** (a) Cascade-type Rydberg electromagnetically induced transparency (EIT) scheme in $^{85}$Rb. A probe field ( $\Omega_P$ ) and a coupling field ( $\Omega_C$ ) establish the EIT resonance, while a microwave field couples adjacent Rydberg states. A strong local oscillator (LO, $\Omega_{LO}$ ) enables superheterodyne detection by down-converting a weak signal (SIG, $\Omega_{SIG}$ ) to an intermediate-frequency modulation. (b) Experimental setup of the Rydberg electrometer with a room-temperature $^{85}$Rb vapor cell; beam displacer (BD), dichroic mirrors (DM), balanced photodetector (BPD) and three-axis compensation coils.

## 3. Effect of magnetic-field compensation on the Rydberg-EIT response

To investigate the impact of magnetic-field on the Rydberg-EIT resonance, we measured the EIT spectra with and without active magnetic-field compensation, as shown in Fig. 2. During this measurement, the coupling laser power was fixed at 30 mW, and all other experimental parameters were kept constant. Figure 2(a) shows the measured EIT transmission spectra with and without magnetic-field compensation. Without compensation (blue), the EIT resonance exhibits a reduced peak amplitude and a broader linewidth. With compensation (red), the EIT signal is significantly enhanced, showing increased amplitude and a narrower linewidth. Quantitatively, the EIT peak amplitude increases by approximately 18%, while the linewidth is reduced from 1.9 MHz to 1.6 MHz, corresponding to a 16% narrowing. These results indicate that environmental magnetic fields introduce additional dephasing and spectral broadening, which are effectively suppressed by magnetic-field compensation.

To analyze this effect quantitatively, we extract the EIT linewidth and peak amplitude under both conditions as a function of probe power, as shown in Fig. 2(b). The blue data points represent the spectrum under the ambient background field, where residual Zeeman splitting causes inhomogeneous broadening of the atomic resonance. Under active compensation (red data points), this broadening is significantly suppressed, yielding a much sharper and more well-defined spectral feature. As shown in Fig. 2(b), the EIT linewidth increases systematically in the absence of magnetic-field compensation. With compensation, the linewidth is reduced and approaches the transit-time broadening limit, demonstrating that magnetic-field-induced broadening is effectively suppressed. In addition to linewidth narrowing, magnetic-field compensation significantly enhances the EIT signal amplitude. The increase in peak transmission reflects improved coherence of the dark-state resonance and reduced spectral averaging due to Zeeman shifts. These results confirm that magnetic-field fluctuations not only broaden the resonance but also degrade the achievable signal contrast.

From the perspective of electric-field sensing, these refinements lead to a direct gain in field responsivity. The electric-field sensitivity is enhanced by a steeper EIT spectral slope, which is fundamentally determined by both the linewidth and the signal amplitude. By simultaneously reducing the linewidth and increasing the peak amplitude, magnetic-field compensation substantially increases the spectral slope at the optimal operating point. These observations highlight that suppressing magnetic-field is essential for accessing the quantum-noise-limited regime.

To theoretically validate the observed probe power dependence and determine the optimal operating conditions, we modeled the macroscopic optical response of the four-level cascade-type EIT system of Fig. 1(a). By explicitly incorporating the residual Doppler shift from the counter-propagating beams and integrating the atomic coherence over a Maxwell-Boltzmann velocity distribution, we applied the Beer-Lambert law to obtain the transmitted signal. This numerical approach successfully reproduced the nonlinear saturation and power broadening effects shown in Fig. 2 (b). These calculations were performed with a fixed coupling Rabi frequency of $\Omega_c / 2\pi = 3.0\,\mathrm{MHz}$ , while the probe power was systematically varied (See Supplementary Note 1 for numerical modeling of probe power dependence on the Rydberg-EIT signal). Our theoretical results (red solid curve) closely agree with the experimental data. Furthermore, applying the same model to the uncompensated data yields an excellent fit (blue curve). This consistent agreement highlights that active B-field compensation is a prerequisite for resolving fine spectral features and achieving the ultimate sensitivity limit.

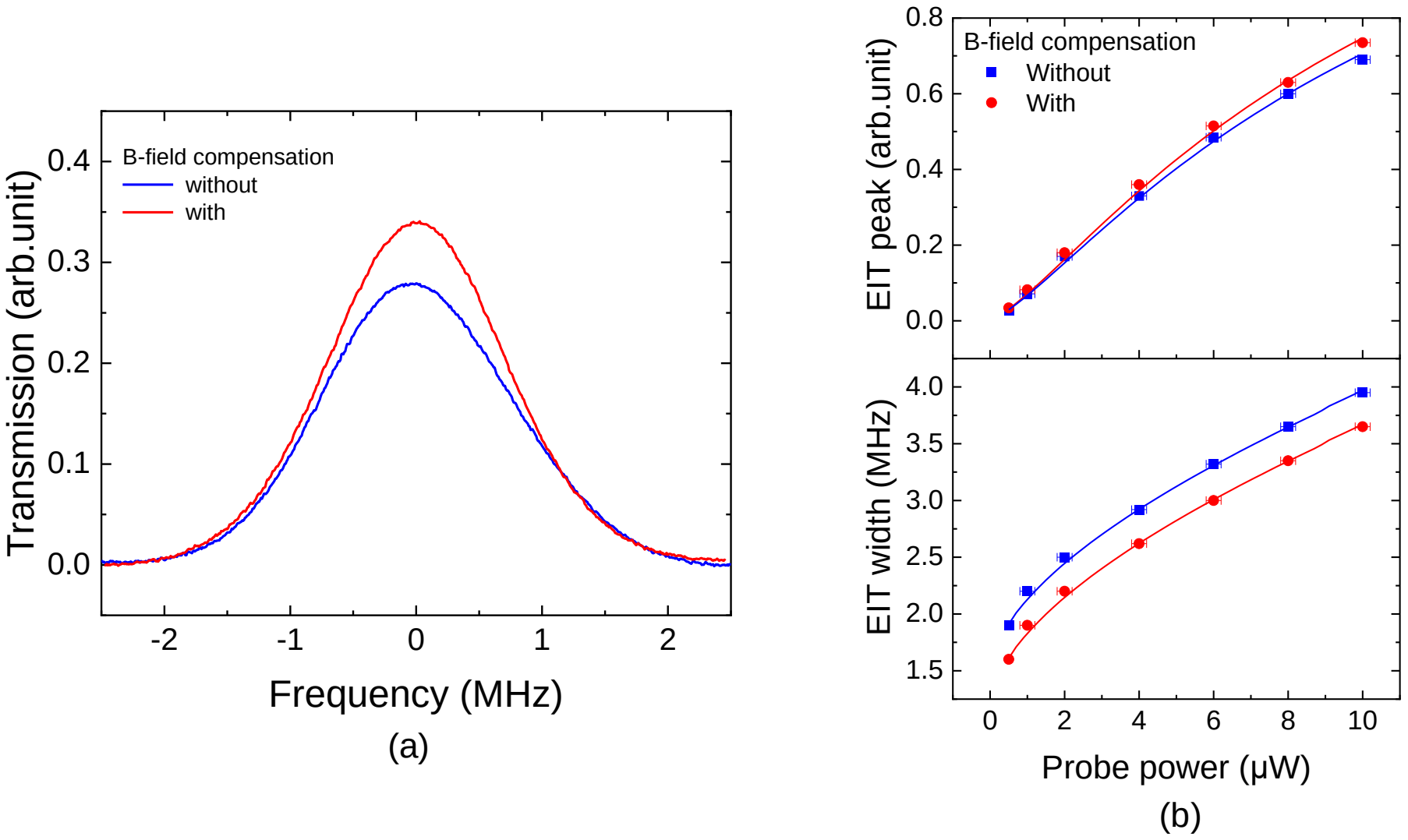

**Fig. 2. Effect of magnetic-field compensation on the Rydberg-EIT response.** (a) Measured EIT transmission spectra with (red) and without (blue) magnetic-field compensation, showing linewidth narrowing and increased peak transmission. (b) Dependence of the EIT peak amplitude (top) and linewidth (bottom) on the probe laser power, compared with (red) and without (blue) the compensation. The solid lines represent theoretical fits to the experimental data.

## 4. Atomic superheterodyne response and temporal signal extraction

We measured the Rydberg-EIT response in the presence of a strong LO field and a weak SIG field. The probe power was fixed at 6 μW, and the LO frequency was set to 37.07104 GHz. A SIG field was applied with a frequency offset of 40 kHz from the LO. Fig. 3(a) shows the measured AT splitting spectra. The blue trace corresponds to the LO-only case, where a symmetric AT splitting is observed, indicating coherent dressing of the Rydberg transition. When the weak SIG is added (red trace), the overall spectral shape remains largely unchanged, with only a slight perturbation near the center of the splitting. This is because the steady-state frequency-domain measurement time-averages the response, meaning that the beating between the LO and SIG is not directly resolved in the spectrum.

To reveal the heterodyne signal, we analyzed the time-domain response at the optimal operating point indicated by the dashed region in Fig. 3(a). The corresponding temporal signals are shown in Fig. 3(b). In the LO-only case (blue trace), the detected signal remains stable over time, confirming the absence of any periodic oscillation. In contrast, when the weak SIG is present (red trace), a clear sinusoidal pattern is readily observed. This oscillation corresponds to the beating between the LO and SIG at 40 kHz, consistent with the applied frequency offset. These results highlight the key feature of atomic superheterodyne detection: the weak SIG is efficiently converted into a large and detectable modulation in the time domain. The LO acts as a gain mechanism that linearizes the response and enhances the detectability of the weak SIG. This behavior confirms that the sensitivity of the Rydberg electrometer is maximized at the point of maximum slope, where small field variations are converted into large temporal modulations.

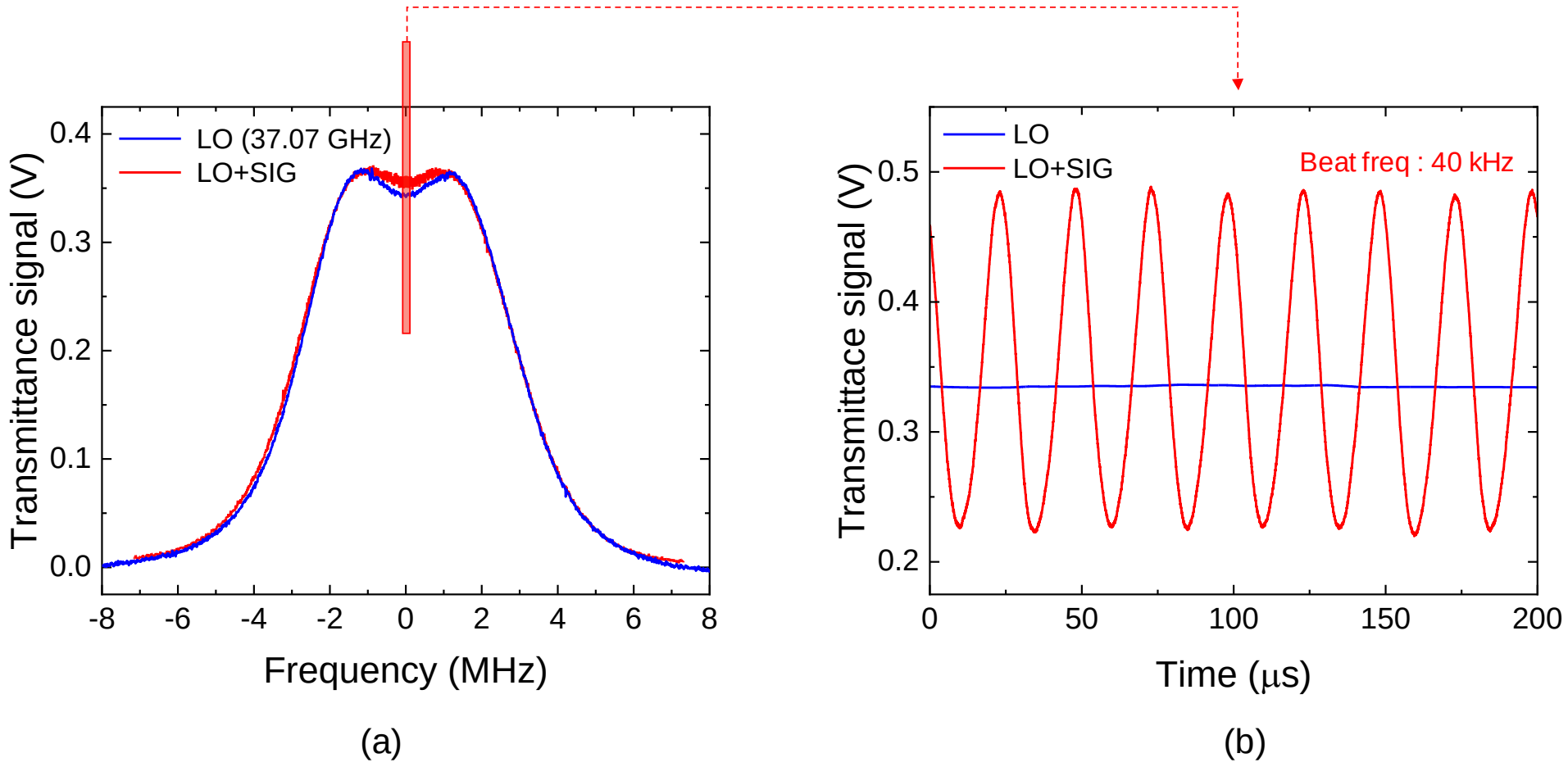

**Fig. 3. Atomic superheterodyne response of the Rydberg-EIT system.** (a) Autler-Townes splitting with a strong LO (blue) and with an additional weak signal field (red). (b) Time-domain probe transmission measured at the operating point in (a), showing a clear oscillation at the LO–signal frequency difference (40 kHz), while the LO-only signal remains stable.

## 5. Electric-field sensitivity and shot-noise-limited operation

In the present work, the term "shot-noise-limited" specifically refers to the probe-photon-shot-noise floor of the optical readout, rather than to the atom projection noise limit of an ideal lossless quantum sensor. This distinction is essential for Rydberg-EIT electrometry, where the experimentally relevant bound is set by the combination of EIT slope and optical shot noise.

In this context, the observed heterodyne beat signal amplitude, $A(\delta)$, is expressed as $A(\delta)=|\kappa_0|\Omega_{SIG}$, where $|\kappa_0|$ and $\Omega_{SIG}$ denote the atomic response and the signal Rabi frequency, respectively. By incorporating the fundamental relationship $\Omega_{SIG}=\mu_{34}E_{SIG}/\hbar$ and adjusting for the root-mean-square (RMS) value, we find that the sensed electric field is independent of extrinsic classical factors such as antenna gain or cable attenuation. Instead, it

relies on the electric dipole moment ( $\mu_{34}$ ) of the Rydberg transition ($40D_{5/2}$ - $39F_{7/2}$), thereby ensuring an inherently SI-traceable measurement capability without the need for additional calibration procedures [6, 8, 13-18]. The signal microwave electric field $E_S$ can be expressed as follows

$$E_S = \frac{1}{\sqrt{2}} \frac{\hbar}{\mu_{34}} \frac{A(\delta)}{|\kappa_0|}. \tag{1}$$

As described in Eq. (1), the microwave-induced Rydberg energy shift is manifested as a modulation of the probe transmission through atomic coherence. In this process, the spectral slope $|\kappa_0|$ serves as a bridge that determines how efficiently the atomic response is converted into an optical signal.

Following the noise analysis established for atomic heterodyne detection [5], the minimum detectable field is inversely proportional to this spectral slope; a steeper slope allows for the detection of smaller signals. The ultimate detection limit is then determined by the level of optical shot noise relative to the slope. Based on this physical framework, the final shot-noise-limited sensitivity $S$ is expressed as shown in Eq. (2):

$$S = \frac{\sqrt{2T_p \hbar^3 \omega_p / P_p}}{\mu_{34} |\kappa_0|}, \tag{2}$$

where $P_p$ is the incident probe power, $\omega_P$ is the probe angular frequency, and $T_P$ is the transmitted probe power fraction. Equation (2) clearly demonstrates that the PSN limit is determined by two competing factors: the probe photon flux and the responsivity $|\kappa_0|$ . Increasing the probe power reduces the relative photon shot noise, but simultaneously, excessive probe power broadens the EIT resonance and consequently decreases the spectral slope. The optimal operating point is therefore obtained by balancing these two effects.

We quantified the electric-field sensitivity and identified the probe-power regime enabling near–shot-noise-limited operation. To evaluate the sensitivity of the Rydberg electrometer, we compared the measured fields with the values derived from the standard antenna method [62], while varying the probe power from 0.5 to 10 μW. The theoretical microwave electric field amplitude, $E_{cal}$ , at a far-field distance r from the transmitting antenna is determined by:

$$E_{cal} = \sqrt{\frac{Z_0 G \alpha_l P_{MW}}{4\pi r^2}}, \tag{3}$$

where $Z_0$ is the free-space impedance, $G$ is the antenna gain, $P_{MW}$ is the input power, and $\alpha_l$ is the system loss factor.

Figure 4(a) presents a comparative analysis of the electric field strengths obtained via both independent methods. The red data points correspond to measurements obtained using the conventional AT-splitting method and the blue data points represent the results from atomic superheterodyne detection. We calculated field $E_{cal}$ (red solid line) as a primary reference from the AT-splitting data (red data points) by directly measuring the spectral splitting in the strong-field regime.

As shown in Fig. 4(a), the measured electric field $E_S$ (blue data points) exhibits excellent linear agreement with both the empirical AT-splitting results and the theoretical $E_{cal}$ over a wide dynamic range. This consistent agreement confirms that our system can accurately determine absolute electric field strengths traceable to SI units without extrinsic calibration.

The sensitivity was then determined under optimized superheterodyne operating conditions for each probe power. Figure 4(a) shows the measured electric field $E_S$ of SIG compared with the calculated electric field $E_{cal}$ derived from the standard antenna method at a probe power of 6 μW. Although both methods exhibit a linear response to the applied electric field, the superheterodyne method demonstrates improved resolution at low field amplitudes, signifying enhanced sensitivity.

Prior to evaluating the field responsivity, we optimized the local oscillator (LO) power to maximize the normalized beating amplitude $\mathrm{A}/\mathrm{A}_{\max}$ (See the Supplementary Note 2 for optimization of the LO power for maximum sensitivity). With the LO power fixed at this optimized level, we extracted the slope of the linear response for different probe powers, as shown in Fig. 4(b). The slope $\kappa_0$ increases with probe power, reflecting an improved signal-to-noise ratio and enhanced optical readout efficiency. The probe power increasing is accompanied by increased power broadening of Rydberg-EIT, leading to a trade-off between signal gain and spectral linewidth.

Figure 4(c) presents the minimum detectable beating amplitude $\mathrm{A}_{\min}$, extracted from the data in Fig. 4(a) as a function of probe power. The value of $\mathrm{A}_{\min}$ decreases with increasing probe power at low power levels from 2 μW to 6 μW, indicating improved sensitivity. At higher probe powers of above 8 μW, however, the $\mathrm{A}_{\min}$ significantly increases due to EIT spectral broadening and increased PSN.

The resulting minimum detectable electric-field intensity $E_{\min}$ can be determined by the relationship between the field responsivity $\kappa_0$ from Fig. 4(b) and the $\mathrm{A}_{\min}$ of Fig. 4(c), accounting for the relevant system constants. Figure 4(d) shows $E_{\min}$ as a function of probe power. We can see the best RF electric-field sensitivity of 12.5(8) nV $\mathrm{cm}^{-1}$ $\mathrm{Hz}^{-1/2}$ for a 1 s integration time at the optimal probe power of 6 μW. To the best of our knowledge, this sensitivity represents the highest experimentally demonstrated sensitivity achieved in vapor-cell-based Rydberg-EIT electrometry.

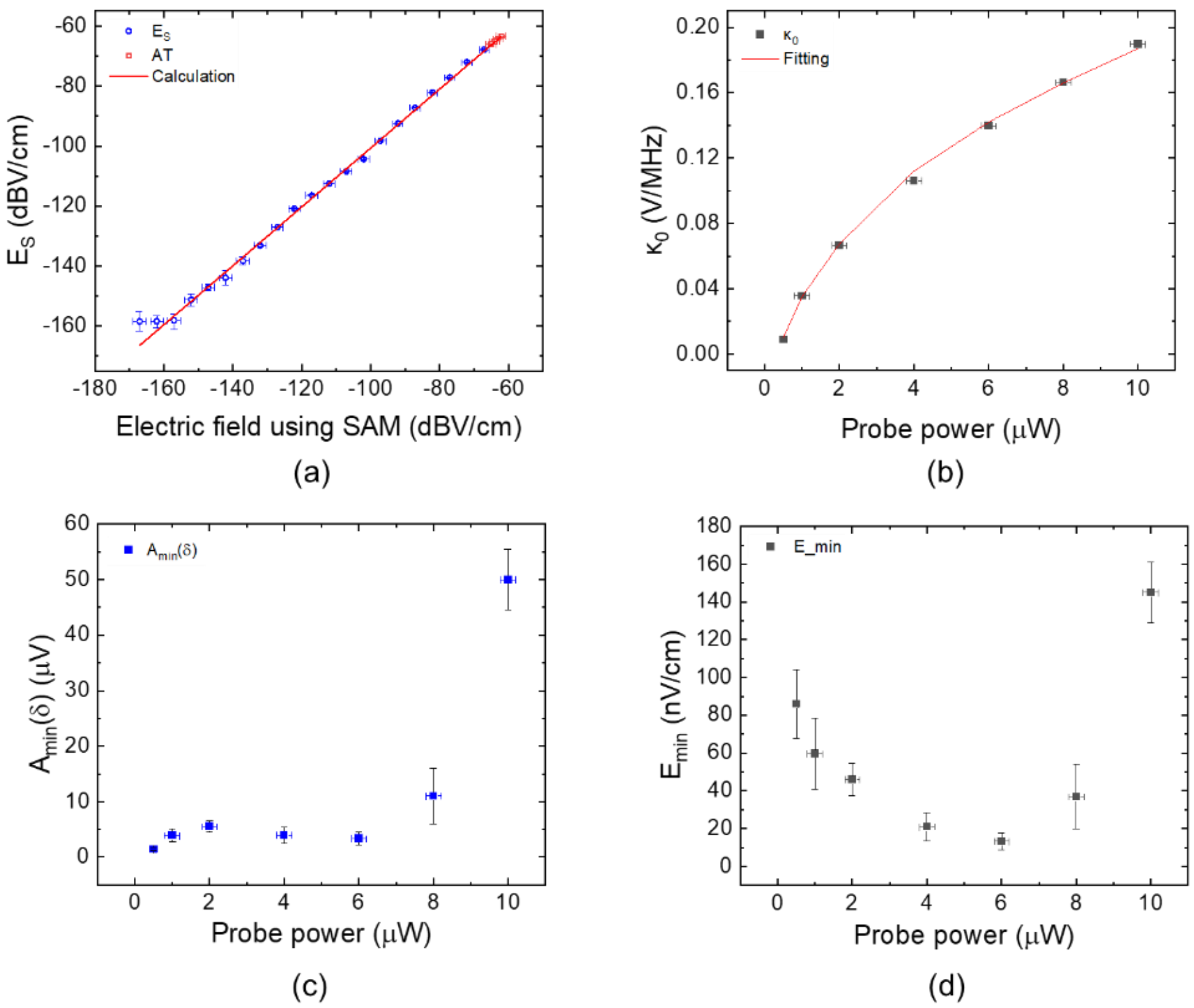

**Fig. 4. Electric-field sensitivity and probe-power optimization.** (a) Measured electric-field response using AT splitting (red) and superheterodyne detection (blue), showing linear agreement with the calculated field. (b) Extracted slope of the response as a function of probe power. (c) Minimum detectable signal amplitude. (d) Electric-field sensitivity, exhibiting an optimum at 6 μW near the photon shot-noise limit.

To further quantify the fundamental sensitivity limit of the Rydberg electrometer, we evaluated the PSN limit under realistic experimental conditions and compared it with the measured sensitivity. Figure 5(a) shows the calculated number of atoms participating in the light–atom interaction as a function of vapor-cell temperature. The calculation was performed using the experimental geometry, with a cell length of 50 mm and a beam diameter of 0.15 mm. As the temperature increases, the atomic density rises, leading to a rapid increase in the effective atom number. A discontinuity appears near 39 °C due to a phase transition in the vapor pressure model [63]. Based on this atom number, we calculated the PSN-limited electric-field sensitivity, shown as the red curve in Fig. 5(a). The PSN limit exhibits a minimum around 33 °C, where the balance between atomic density and optical response is optimized. In the present experiment, however, our vapor cell was operated at room temperature (~20 °C), where the atom number is lower and the corresponding PSN limit is slightly higher than its optimal value.

Figure 5(b) compares the experimentally measured electric-field sensitivity (blue circles) with the calculated PSN limit (red dashed line), as a function of probe power under the experimental temperature condition. The measured sensitivity follows the same overall trend as the calculated shot-noise limit, confirming that the dominant noise source is consistent with probe photon shot noise. In particular, at a probe power of 6 μW, the measured sensitivity approaches the calculated PSN limit of 11.2 nV cm⁻¹ Hz⁻¹ᐟ².

These results highlight the importance of probe-power optimization in Rydberg electrometry. The experimental sensitivity reaches its minimum near the predicted optimum, demonstrating that the close agreement between theory and experiment validates our framework. Consequently, the present compact vapor-cell electrometer operates in the near–shot-noise-limited regime.

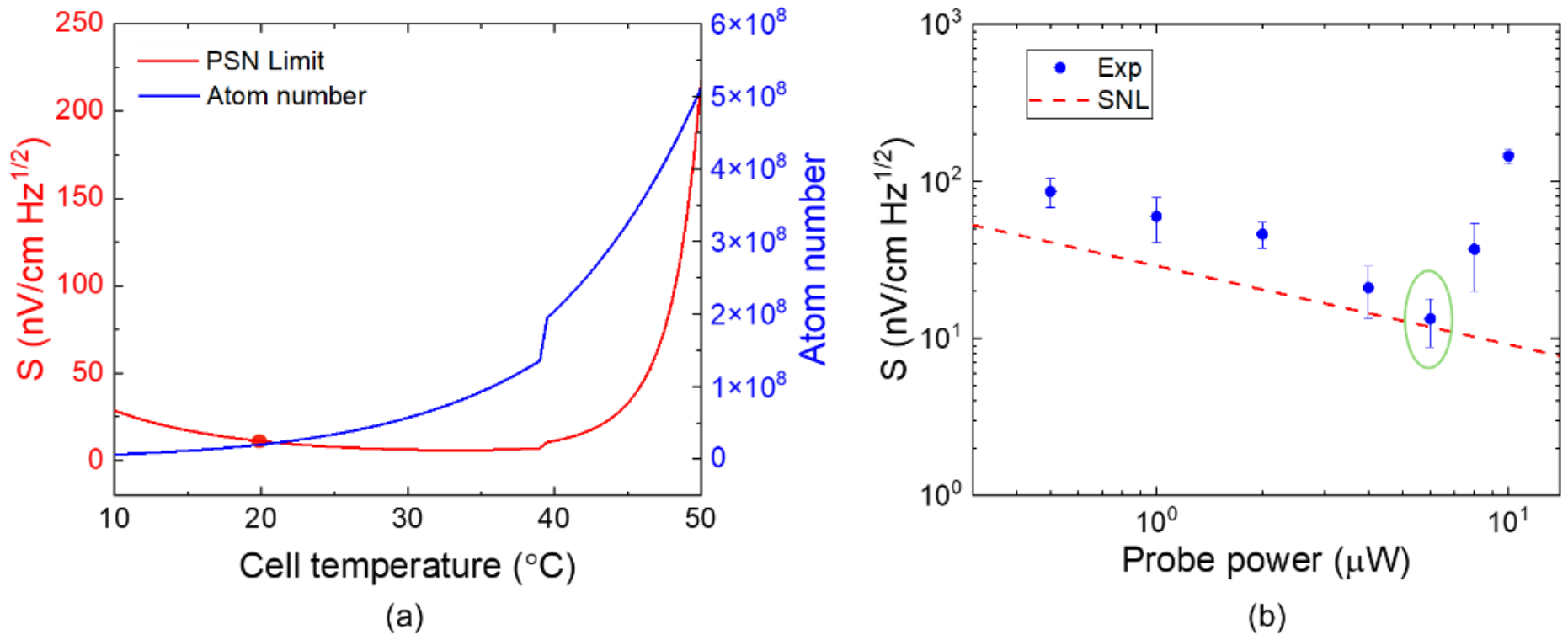

**Fig. 5. Comparison with the photon shot-noise (PSN) limit.** (a) Calculated atom number (blue) and PSN-limited sensitivity (red) as a function of vapor-cell temperature. (b) Measured sensitivity (blue) and calculated PSN limit (red dashed line) versus probe power, showing agreement near the optimal operating point. The sensitivity approaches the shot-noise limit near 6 μW (green circle).

To contextualize the performance of our optimized sensor, Table 1 summarizes the historical progress of sensitivities reported for Rydberg atom-based microwave electrometry. By systematically mitigating inhomogeneous Zeeman broadening and identifying the optimal probe and LO powers, our superheterodyne system achieved a highly competitive sensitivity of 12.5(8) nV cm⁻¹ Hz⁻¹ᐟ² in a room-temperature vapor cell. As shown in the table, this result significantly outperforms earlier homodyne and heterodyne techniques and closely approaches the ultimate sensitivity limits recently demonstrated by complex cold-atom systems.

**Table. 1. Comparison of state-of-the-art Rydberg-atom microwave electrometry sensitivities**. Summary of reported noise-equivalent electric fields across various detection schemes and experimental platforms.

| **Atomic system** | **Method** | **Frequency (GHz)** | **Sensitivity, *S* (nV cm⁻¹ Hz⁻¹ᐟ²)** | **Ref** |
|---|---|---|---|---|
| **Rb vapor cell** | Probe transmission | 14.2 | 30,000 | [1] |

| **Cs vapor cell** | Homodyne | 5.0 | 5,000 | [28] |
|---|---|---|---|---|
| **Cs vapor cell** | Frequency modulated spectroscopy | 5.0 | 3,000 | [58] |
| **Cs vapor cell** | Heterodyne | 19.6 | 790 | [21] |
| **Cs vapor cell** | Superheterodyne | 6.9 | 55 | [6] |
| **Rb vapor cell** | Repumping | 17.0 | 30 | [47] |
| **Rb vapor cell** | Optimization | 10.7 | 35.75 | [48] |
| **Cs vapor cell** | Self-heterodyne | 19.5 | 23 | [31] |
| **Cold Rb** | Superheterodyne | 6.9 | 10 | [5] |
| Rb vapor cell | Superheterodyne with optimization | 37.1 | 12.5 | This work |

## 6. Conclusion

We have experimentally demonstrated vapor-cell-based Rydberg-EIT electrometry operating near the PSN limit with the highest sensitivity reported to date. By systematically suppressing residual magnetic-field and laser noise, we optimized the EIT linewidth, which determines the field responsivity. Our results clarify the interplay between linewidth, optical power, and superheterodyne configurations, and show that the sensitivity of Rydberg sensors is governed by a balance between spectral slope and photon shot noise rather than by a single technical parameter. By optimizing the parameters of superheterodyne detection, we demonstrated operation in a regime where the optical readout approaches its fundamental limit, yielding a highly competitive sensitivity of 12.5(8) nV $cm^{-1}$ $Hz^{-1/2}$. Our approach provides a unified framework for understanding and optimizing the performance of Rydberg-EIT electrometers, closely approaching the capabilities of cold-atom systems [5]. More fundamentally, this work establishes a quantitative framework linking spectral responsivity, linewidth engineering, and optical shot noise in Rydberg-EIT electrometry. Our results therefore identify the key physical requirements for realizing quantum-limited electric-field sensing in thermal atomic platforms.

**Funding.** This work was supported by the Korea Research Institute for Defense Technology Planning and Advancement (KRIT) through a grant funded by the Defense Acquisition Program Administration (DAPA) (KRIT-CT-23-031), the Institute for Information and Communications Technology Promotion (RS-2024-00396999, IITP-2026-RS-2020-II201606, RS-2022-II221029, and RS-2025-25464481) and the National Research Foundation of Korea (RS-2023-00283146 and RS-2025-25454381).

**Disclosures.** The authors declare no conflicts of interest.

**Data availability.** Data underlying the results presented in this paper are not publicly available at this time but may be obtained from the authors upon reasonable request.

**Supplemental document.** See the Supplementary Materials for detailed descriptions: Numerical modeling of probe power dependence on Rydberg-EIT and optimization of the local oscillator (LO) power for maximum sensitivity.